\documentclass[12pt]{article}
\usepackage{amssymb}

\usepackage{epsfig}

\def\beq{\begin{equation}}
\def\eeq{\end{equation}}

\def\gev{\, {\rm GeV}}

\def\bea{\begin{eqnarray}}
\def\eea{\end{eqnarray}}
\def\beq{\begin{equation}}
\def\eeq{\end{equation}}
\def\mtilde{\tilde m}
\newcommand{\gsim}{\lower.7ex\hbox{$\;\stackrel{\textstyle>}{\sim}\;$}}
\newcommand{\lsim}{\lower.7ex\hbox{$\;\stackrel{\textstyle<}{\sim}\;$}}

\begin{document}


\begin{titlepage}
\noindent
\begin{flushright}
ANL-HEP-PR-03-010 \\
MCTP-03-04\\
\end{flushright}
\vspace{1cm}

\begin{center}
  \begin{large}
    \begin{bf}
Abelian $D$-terms and the superpartner \\
spectrum of anomaly-mediated supersymmetry breaking 
    \end{bf}
  \end{large}
\end{center}
\vspace{0.2cm}
\begin{center}
Brandon Murakami${}^{a}$ and James D. Wells${}^{b}$ \\
  \vspace{0.2cm}
  \begin{it}
${}^{(a)}$ \
HEP Division, Argonne National Lab, 9700 Cass Ave, Argonne, IL 60439 \\
${}^{(b)}$ 
MCTP, Department of Physics, University of Michigan, Ann Arbor, MI 48109 

  \end{it}

\end{center}

\begin{abstract}

We address the tachyonic slepton problem of anomaly mediated
supersymmetry breaking using abelian $D$-terms.  We demonstrate that
the most general extra $U(1)$ symmetry that does not disrupt gauge
coupling unification has a large set of possible charges that solves
the problem.  It is shown that previous studies in this direction that
added both an extra hypercharge $D$-term and another $D$-term induced
by $B-L$ symmetry (or similar) can be mapped into a single $D$-term of
the general ancillary $U(1)_{\rm a}$.  The $U(1)_{\rm a}$ formalism
enables identifying the sign of squark mass corrections which leads to
an upper bound of the entire superpartner spectrum given
knowledge of just one superpartner mass.

\end{abstract}

\vspace{1cm}

\begin{flushleft}
hep-ph/0302209 \\
February 2003
\end{flushleft}

\end{titlepage}


The pure anomaly-mediated supersymmetry breaking (AMSB) spectrum includes
negative squared masses for all sleptons, which inflicts an unwanted
Higgs mechanism upon electromagnetism. If it were not for this
unfortunate feature, AMSB~\cite{Randall:1998uk}-\cite{Bagger:1999rd}
would be an ideal candidate for the standard model of supersymmetry
breaking transmission.  It has the virtues of being ultraviolet (UV)
insensitive, compatible with flavor changing neutral current
constraints, and requiring modest model building from the low-energy
perspective.  Of course, the viability of any AMSB model with two-loop
suppressed scalar masses requires significant cooperation from the the
K\"ahler potential, superpotential, and gauge kinetic
function~\cite{Randall:1998uk}, which could be accomplished by either
geometric means of an extra dimensional brane world supersymmetry
breaking or a purely four dimensional conformal hidden
sector~\cite{Luty:2001zv}.

Salubrious implications to long-standing issues of supersymmetry are
not the only alluring aspects of AMSB.  AMSB contributions to the
masses are not optional -- they are present in all broken supergravity
theories. We wish, therefore, to make a few contributions to the
discussion of AMSB, and in particular regarding the $D$-term solution
of the tachyonic slepton problem due to an extra $U(1)$ symmetry in the
theory.  Our considerations have led us to a general ``ancillary
$U(1)_{\rm a}$'', to be defined below, which nicely parametrizes some
of the previous studies in this direction, and enables us to make quick
work of some interesting theorems regarding the superpartner spectrum.  

Recall that the AMSB masses~\cite{Randall:1998uk,Giudice:1998xp}, which are
invariant expressions under renormalization group evolution (``RG
invariant''), are in the minimal supersymmetric standard model (MSSM) given by
\bea
M_\lambda & = & \frac{\beta_g}{g}m_{3/2} \label{gauginomasses}\\
m^2_{\tilde Q} & = & -\frac{1}{4}\left( 
\frac{\partial \gamma}{\partial g}\beta_g +\frac{\partial \gamma}{\partial y}
\beta_y\right) m_{3/2}^2 \label{scalarmasses} \\
A_y &= & -\frac{\beta_y}{y}m_{3/2} , \label{trilinears}
\eea
where $m_{3/2}$ is the gravitino mass.  Since the masses are one-loop
supressed compared to the gravitino mass, it will be convenient to
define the mass scale $\mtilde \equiv m_{3/2}/16\pi^2$.  Thus,
$\mtilde$ is a more direct estimator of the superpartner masses than
$m_{3/2}$. Expressions for the pure AMSB masses in terms of
$\mtilde=m_{3/2}/16\pi^2$, gauge couplings and Yukawa couplings can be
found in many places in the literature (see, e.g., Eqs. (44)-(59)
of~\cite{Gherghetta:1999sw}).

Application of Eq.~(\ref{scalarmasses}) to sleptons shows that their
mass-squared is negative. Many groups have proposed ideas to solve this
tachyonic slepton
problem~\cite{Pomarol:1999ie},\cite{Katz:1999uw}-\cite{Jack:2003qg}.
Some solutions lose the UV insensitive feature of AMSB, and some
solutions preserve it.  We wish to keep UV insensitivity, and retain
gauge coupling unification, which is one of the more powerful hints
that supersymmetry is a necessary element of the full theory of nature.
The combination of these requirements leads us to consider low-scale
$D$-terms from broken exotic gauge symmetries.

There are many considerations in employing $D$-term slepton solutions.
This list includes the origin of the $D$-term VEV, the crossing of
thresholds, the charge assignments of the $U(1)$ symmetry, proper
electroweak symmetry breaking, gauge kinetic mixing with hypercharge,
and gauge boson mass mixing.  Some of these issues can only be answered
with detailed model building. Our purposes here are to prove results on
the superpartner spectrum that will be true for all models subject to
only a few basic axioms.

Our basic hypotheses are that the only low-scale fields that feel SM
charges are the MSSM fields.  Furthermore, the sleptons obtain positive
mass-squareds via $D$-terms of an ancillary abelian symmetry, which is
defined to be a local $U(1)$ symmetry that is anomaly free but whose
charges do not alter standard gauge coupling unification. We assume and
expect that the chiral multiplets that break the $U(1)_{\rm a}$ have no
substantive Yukawa couplings to MSSM fields that would affect the AMSB
spectrum of Eqs.~(\ref{gauginomasses})-(\ref{trilinears}).  Our working
hypotheses imply that Standard Model (SM) particles can be charged
under the additional $U(1)_{\rm a}$, but exotic states needed for
anomaly cancellation cannot be charged under the SM.  $U(1)_{\rm a}$
symmetry breaking will then result in $D$-terms masses for each of the
MSSM states,
\beq
(m^{D_{\rm a}}_i)^2 = Q_{\rm a}^i D_{\rm a}
\label{Qamass}
\eeq
where $Q_{\rm a}^i$ is the $U(1)_{\rm a}$ charge, and $D_{\rm a}$
absorbs the $U(1)$ gauge coupling, and is defined by the above equation.

\begin{table}[t]
\small
\centering
\begin{tabular}{ccccc}
\hline\hline
 & $Q_{\rm a}^i$  & $Y$ & $B-L$ & $J$ \\
    & ($q,r$ free) &  ($q=-1/6$, $r=1$)    
  &  ($q=-1/3$, $r=1$)   &   ($q=-7/3$, $r=3$)  \\ 
\hline
$L$ & $3q$   & $-1/2$  & $-1$  & $-7$ \\
$e$ & $r$    & $1$  & $1$ & $3$ \\
$Q$ &  $-q$           & $1/6$    &  $1/3$      & $7/3$ \\
$u$ & $-2q-r$      & $-2/3$   & $-1/3$  & $5/3$ \\
$d$ & $4q+r$   & $1/3$  & $-1/3$ & $-19/3$ \\
$H_u$ & $3q+r$ & $1/2$ & $0$ & $-4$ \\
$H_d$ & $-3q-r$ & $-1/2$ & $0$ & $4$ \\
\hline\hline
\end{tabular}
\caption{Table of charges for the ancillary $U(1)_{\rm a}$ symmetry. It
is defined to be a symmetry that can be made anomaly free without the
need of adding any additional states charged under the SM. It can be
parametrized by two rational constants~\cite{Chamseddine:1995rs}.
Hypercharge, $B-L$, and the $U(1)_J$ symmetry of
Ref.~\cite{Jack:2000cd} are all special cases of $U(1)_{\rm a}$
generated by the appropriate two parameter choice $(q,r)$.}
\label{chargetable}
\end{table}

The SM charges under this additional $U(1)_{\rm a}$ can be classified
by two free parameters~\cite{Chamseddine:1995rs} (see
Table~\ref{chargetable}). We know that the slepton masses must have a
large, positive contribution from the $D_{\rm a}$ term, so we know that
$r>0$ and $q>0$. This gives us freedom to normalize the charge of the
right-handed electron chiral superfield to be $r=1$, and let the
overall size of the mass correction be controlled by the $D$-term vev.
The remaining charges are then dependent on only one variable $q$:
\beq
\begin{array}{c}
\begin{array}{cccc}
Q_e=+1, & Q_L=3q, & Q_{H_u}= 3q+1, & Q_{H_d} = -(3q+1),
\end{array}\\
\begin{array}{ccc}
Q_Q = -q, & Q_u = -(2q+1), & Q_d = 4q+1 .
\end{array}
\end{array}
\label{mattercharges}
\eeq
We do not address here the precise charges
of the exotic SM singlet states $s_i$,  except to note that their
$U(1)_{\rm a}$ charges must satisfy
\beq
\sum s_i = -3(6q+r) ~~~ {\rm and}~~~ \sum s_i^3= -3(6q+r)^3.
\eeq
Although large Yukawa couplings are generically necessary to break the
$U(1)_{\rm a}$, renormalizable operators do not couple the MSSM matter
and the SM singlets charged under the $U(1)_{\rm a}$.  
Therefore, there will be no additional AMSB
contribution of the Yukawa variety. (Pure singlets, which we eschew, 
would add Yukawa complications  when interacting with $H_uH_d$.)
Solutions for $s_i$ charges for a
particular set of fields and symmetries is a model building question
that we will not discuss here, except to say that some simple cases were
analyzed in~\cite{Chamseddine:1995rs,Jack:2000cd}. Consistent with our
present goals, we will only discuss results that must hold for all
possible model choices.

We consider the general case of a single arbitrary $U(1)_{\rm a}$,
determine what values of $q$ and $D_{\rm a}$ are needed to solve the
slepton problem, and then determine what predictions follow.  To more
easily keep track of scale factors, we will define $D_{\rm a}=\eta
\mtilde^2$. To generate the entire superpartner spectrum we need only
specify 
\beq
\mtilde,\, \tan\beta,\,\eta,\, q,\, {\rm and}\, b_{\rm a}g_{\rm a}^4
~~~({\rm input~parameters})
\eeq
where $b_{\rm a}g_{\rm a}^4$ characterizes the anomaly mediated 
contribution due to the $U(1)_{\rm a}$ and 
is clarified in Eq.~(\ref{deltagamma}) below.

We can then compute the $D$-term contribution to the superpartner
masses:
\bea
(m_L^{D_{\rm a}})^2 & = & 3q\eta\tilde m^2 \nonumber\\
(m_e^{D_{\rm a}})^2 & = &\eta\mtilde^2 \nonumber\\
(m_Q^{D_{\rm a}})^2 & = & -q\eta\tilde m^2 \nonumber\\
(m_u^{D_{\rm a}})^2 & = & (-2q-1)\eta \mtilde^2 \nonumber\\
(m_d^{D_{\rm a}})^2 & = & (4q+1)\eta \mtilde^2 \nonumber\\
(m_{H_u}^{D_{\rm a}})^2 & = & (3q+1)\eta\mtilde^2 \nonumber\\
(m_{H_d}^{D_{\rm a}})^2 & = &-(3q+1)\eta\mtilde^2 .
\label{Dtermmasses}
\eea
Additionally, a light, propagating $Z'$ associated with this $D$-term
(which is not necessarily required~\cite{Jack:2000cd,
Arkani-Hamed:2000xj}) would alter the anomalous dimensions of the
matter fields by $\Delta\gamma_i = 4(Q_{\rm a}^i)^2 g_{\rm a}^2$
resulting in an AMSB contribution to the soft masses:
\bea
(m_L^{\Delta\gamma})^2 &=& -2 (3q)^2 b_{\rm a} g_{\rm a}^4 \mtilde^2 \nonumber\\
(m_e^{\Delta\gamma})^2 &=& -2 b_{\rm a} g_{\rm a}^4 \mtilde^2 \nonumber\\
(m_Q^{\Delta\gamma})^2 &=& -2 q^2 b_{\rm a} g_{\rm a}^4 \mtilde^2 \nonumber\\
(m_u^{\Delta\gamma})^2 &=& -2 (2q+1)^2 b_{\rm a} g_{\rm a}^4 \mtilde^2 \nonumber\\
(m_d^{\Delta\gamma})^2 &=& -2 (4q+1)^2 b_{\rm a} g_{\rm a}^4 \mtilde^2 \nonumber\\
(m_{H_u}^{\Delta\gamma})^2 &=& -2 (3q+1)^2 b_{\rm a} g_{\rm a}^4 \mtilde^2 \nonumber\\
(m_{H_d}^{\Delta\gamma})^2 &=& -2 (3q+1)^2 b_{\rm a} g_{\rm a}^4 \mtilde^2 .
\label{deltagamma}
\eea
where $b_{\rm a}$ is the beta function coefficient of $U(1)_{\rm a}$
($dg_a/dt=b_ag_a^3/16\pi^2$).
The effect of this contribution is to drive the sleptons more negative 
in squared mass in the infrared. 

We can immediately construct a few theorems about the spectrum that
arise from the ansatz of ancillary $U(1)$ symmetry.  Since contributions
must be positive for $m_L^2$, we know that $q>0$ is necessary. Therefore,  
by mere inspection of Eqs.~(\ref{Dtermmasses})-(\ref{deltagamma}),
we determine that additional D-term mass contributions to the squarks are 
always of well-defined sign:
\begin{eqnarray}
(m_Q^{D_{\rm a}})^2 < 0, ~~~~
(m_u^{D_{\rm a}})^2 < 0, ~~~~
(m_d^{D_{\rm a}})^2 > 0 .
\end{eqnarray}

In the event that $U(1)$ breaking is at a low scale and
$(m_i^{\Delta\gamma})^2$ mass contributions are present, we can see
that $\eta$ must necessarily be above $2b_{\rm a}g_{\rm a}^4$ in order
to overcome the additional tachyonic contributions $\Delta\gamma$ 
imposes on the sleptons. Nevertheless, we note that the up-squark
mass still {must be} shifted lower in mass compared to the pure MSSM
AMSB mass ($\Delta m_u^2<0$), and the down-squark mass still must be
shifted higher in mass ($\Delta m_d^2>0$). Only the left-squark doublet
mass can be shifted either up or down in this case depending on the
precise relationship between the parameters.

Determining the mass of the squarks with respect to the pure-AMSB value
is another way to test the theory. Precise measurements of heavy
squarks are difficult at colliders, but it is profitable and
feasible to answer simple binary questions of the form ``Is this squark
mass we just measured heavier or lighter than the pure AMSB mass we
infer for it from the gaugino spectrum?''

We also can state another important implication to the superpartner
spectrum.  Given the intrinsic mass scale $\mtilde=m_{3/2}/16\pi^2$, we
can conclude that 
\begin{equation}
{\rm mass~of~any~squark~or~slepton} < 6\mtilde 
\label{masstheorem}
\end{equation}
under the broad assumptions we have outlined for this study.
Measurement of any one superpartner fixes $\mtilde$, and the above
equation can then be used to cap the maximum value of any other
superpartner in the spectrum.  For example, if we measure the lightest
chargino and determine that it is mostly Wino, we can use the resulting
value of $M_2$ ($M_2\simeq 0.4\tilde m$) to place a limit of about
$15M_2$ for any other superpartner. The bound comes from the very
conservative requirement that all first generation superpartner masses
are positive. The constraints from EWSB symmetry breaking and
experimental limits on third generation mass eigenstates makes the
result even tighter for various values of $q$, $\eta$ and $\tan\beta$
input parameters.

\begin{figure}
\centering
\includegraphics*[width=12cm]{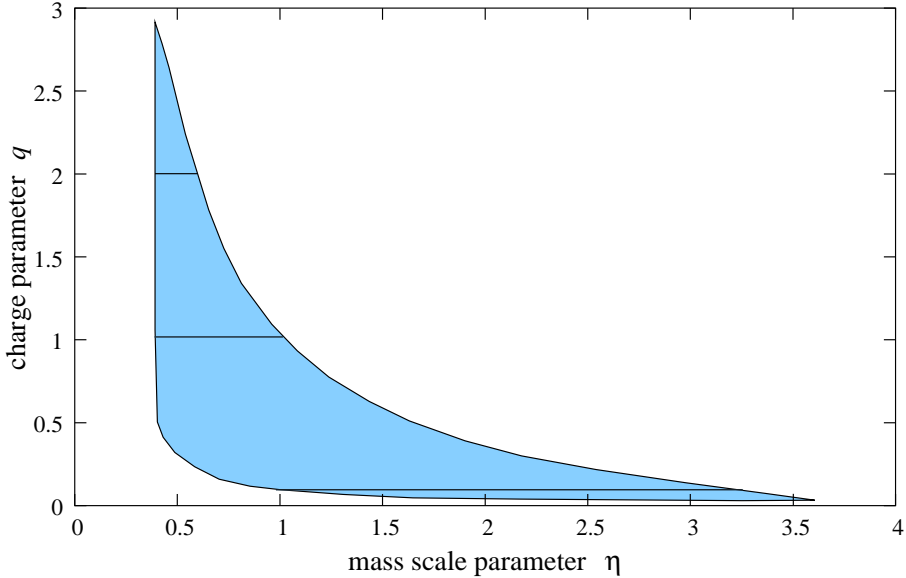}
\caption{The parameter space of AMSB with ancillary $U(1)_{\rm
a}$ is spanned by four parameters, $\mtilde$, $\tan\beta$, $\eta$ and
$q$.  We demonstrate the region of allowed $\eta$ and $q$ that insures
that all sfermions have acceptable masses and allows proper EWSB. For
this example, we have chosen $\mtilde=500\gev$ and $\tan\beta=10$. The
horizontal lines represent the parameter space that compose the
spectrums of Fig.~\ref{spectrums}.}
\label{parameterspace}
\end{figure}

In Fig.~\ref{parameterspace}, we have demonstrated the parameter space
for $q$ and $\eta$ given the other two input parameters fixed at
$\mtilde =500\gev$ and $\tan\beta =10$.  Here (and for Fig.~2, as well),
we have excluded the effects of Eq.~(\ref{deltagamma}).\footnote{This is the
case of Refs.~\cite{Jack:2000cd} and \cite{Arkani-Hamed:2000xj}.} For any
given value of $\eta$, $q$ cannot be too large, otherwise some of the
squark masses will go tachyonic; and $q$ cannot be too low, otherwise
the left-sleptons will remain tachyonic.  Likewise, for any given value
of $q$, $\eta$ cannot be too large, otherwise some of the squarks will
go tachyonic; and $\eta$ cannot be too small, otherwise both right and
left sleptons will remain tachyonic.

If $g_{\rm, a}\sim g_Y$ at the weak
scale, $U(1)_{\rm a}$ would lower all soft masses of scalars charged
under it, resulting in less parameter space.  Alternatively,
one might ponder starting $g_{\rm a}(m_{\rm GUT})$ with the gauge
unification value and running $g_{\rm a}$ to the weak scale. To avoid
committing to a specific model, the beta function coefficient $b_{\rm
a}$ may be approximated by excluding the contributions due to the
unspecified singlet sector, which generally should contribute less.
(Without this assumption, one could find, in principle, parameter space
allowing implausibly large values of $b_{\rm a}$ in the hundreds.) At
the weak scale, this typically would lead to smaller $g_{\rm
a}(m_Z)<g_Y(m_Z)$, expanding the allowed parameter space all the way
to that of Fig.~\ref{parameterspace} in the limit $g_{\rm a}(m_Z)\to 0$.

\begin{figure}
\centering
\includegraphics*[width=12cm]{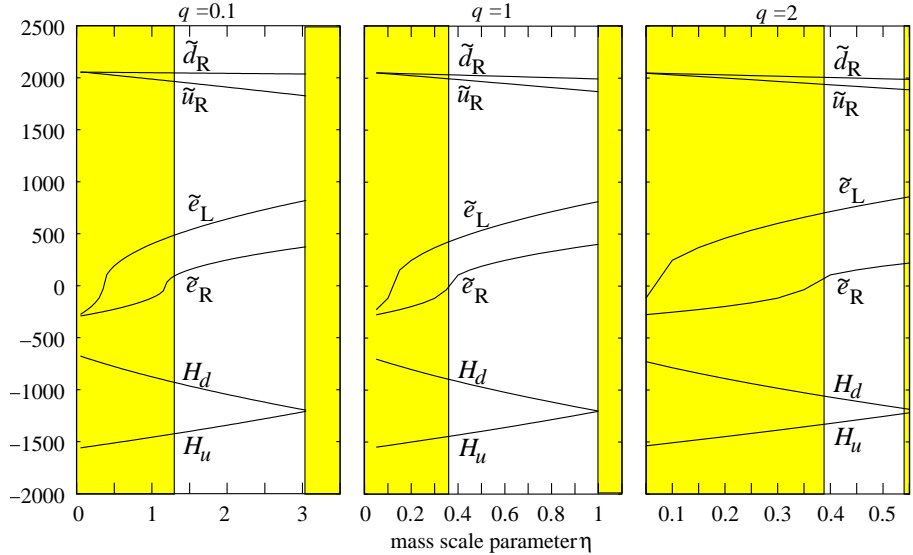}
\caption{The parameter space of AMSB with ancillary $U(1)_{\rm a}$ is
spanned by four parameters, $\mtilde$, $\tan\beta$, $\eta$ and $q$.  We
demonstrate how the first generation scalar soft masses (in GeV) scale
with $\eta$ for three different choices of $q$. For this example, we
have chosen $\mtilde=500\gev$ and $\tan\beta=10$.  The shaded region on
the left is excluded by insufficient slepton mass squared.  The shaded
region on the right is excluded by improper electroweak symmetry
breaking.  Notice the entire spectrum is bounded below about
$4\mtilde$, which is an illustration of how a technical analysis of the
spectrum for any given set of inputs usually imply a tighter,
lower-mass spectrum than the most general expression in
(\ref{masstheorem}).  Negative soft mass values are defined as
$-\sqrt{-m_\phi^2}$.}
\label{spectrums}
\end{figure}

In Fig.~\ref{spectrums}, we demonstrate how the spectrum scales with
$\eta$ for the other three parameters fixed $\mtilde =500\gev$,
$\tan\beta=10$, and $q=0.1, 1, 2$.  The $\eta$ values terminate before
any squark gets dangerously close to zero mass.  This is because EWSB
begins to fail.  The up and down Higgs squared masses are driven in
opposite directions due to their charge assignments.  If $m_{H_u}^2$
and $m_{H_d}^2$ are nearly equal, $B\mu$ goes to a small, non-zero,
positive value.  The classical condition that the Higgs potential be
bounded below then fails.


We now wish to compare our use of $D$-terms as an AMSB slepton solution
directly to previous studies.  Our approach is related to the work in
Ref.~\cite{Jack:2000cd, Arkani-Hamed:2000xj}.  Our discussion is
designed to imagine $U(1)_{\rm a}$ freely parameterizing the $D$-terms
of either scenario.  One simplifying feature of the ancillary $U(1)$
assumption is that it allows for the use of a single exotic $D$-term
while maintaining gauge coupling unification.

The study in~\cite{Jack:2000cd} analyzed the single case
$(q,r)=(-7/3,3)$ originating from the condition ${\rm Tr}(YQ)=0$, which
zeroed one-loop corrections to kinetic mixing.  We call this
special case $U(1)_{\rm J}$. However, this case does not solve the slepton
problem by itself, as the mass-squared contribution to the slepton-left
$L$ is opposite sign to that of slepton-right $e$.  It is then
necessary to introduce an additional $D$-term proportional to the
hypercharge of each superpartner, such that a judicious combination of
the two $D$-terms would yield positive mass-squareds for both
slepton-right and slepton-left. Unspecified, additional model building
complexity would be needed to explain the origin of these additional
hypercharge $D$-terms.

In that framework, the soft masses that compose the FI scale and
singlet VEVs presumably originate from AMSB. However, soft mass
corrections from physics beyond the cutoff scale $\Lambda$ of the MSSM
effective theory are cancelled {exactly} to one loop order. Higher
order corrections begin on the order $m_{3/2}^4/\Lambda^2$. And so
without additional dynamics, relevant scales cannot be too high if they
are to be responsible for sufficient $D$-terms.

The authors of Ref.~\cite{Arkani-Hamed:2000xj} consider the $B-L$ gauge
symmetry, which has opposite sign charges for $L$ and $e$ requiring the
use of a second $D$-term.  They use a hypercharge $D$-term established
through an operator of the form $\int d^2\theta \,  {\cal W}_Y{\cal
W}_{B-L}$, which, in essence, allows free parameterization for the
hypercharge $D$-term. As was the case with $U(1)_{\rm J}$ $D$-terms,
$U(1)_{B-L}$ $D$-terms must be augmented by additional $U(1)_Y$
$D$-terms in order to solve the tachyonic slepton problem in this scenario.

We now demonstrate that our simple parametrization of the $D$ terms is
a general parametrization that can describe the entire parameter space
of both Ref.~\cite{Jack:2000cd} and Ref.~\cite{Arkani-Hamed:2000xj}. 
In both of these analyses, there are two $D$-term contributions to the
scalar masses of the squarks and sleptons.  One $D$-term is an
arbitrary additional contribution to the ordinary hypercharge
$D_Y$-term, and the other $D$-term is proportional to the charges of a
special $U(1)$ which is $B-L$ in Ref.~\cite{Arkani-Hamed:2000xj} or J
in Ref.~\cite{Jack:2000cd}:
\begin{eqnarray*}
\Delta m^2_i & = & Y_iD_Y + Q^{B-L}_iD_{B-L},~~~{\rm or} \\
\Delta m^2_i & = & Y_iD_Y + Q^{\rm J}_i D_{\rm J}.
\end{eqnarray*}
Table~1 contains the charges of each superpartner state in terms of
hypercharge, $B-L$, J symmetry, and our general $U(1)_{\rm a}$ symmetry.

We can recast these $D$-term contributions to the mass of sparticle $i$
in terms of our language (Eq.~\ref{Qamass}) by setting $\eta=1$ and
$r=1$, and then identifying $D_{\rm a}$ and $q$ with
\beq
D_{\rm a} = r_YD_Y + r_{\rm X}D_{\rm X},
\label{Damap}
\eeq
and
\beq
q=\frac{q_Y D_Y + q_{\rm X}D_{\rm X}}{r_YD_Y + r_{\rm X}D_{\rm X}}
\label{qmap}
\eeq
where X is a label for either $B-L$ or J, $(q_Y,r_Y)=(-1/6,1)$,
$(q_{B-L},r_{B-L})=(-1/3,1)$, and $(q_{\rm J},r_{\rm J})=(-7/3,3)$.

These simple mappings of the two $D$-terms analyses of
Refs.~\cite{Jack:2000cd}~and~\cite{Arkani-Hamed:2000xj} are made
possible because all $U(1)$'s under discussion are special cases of
$U(1)_{\rm a}$.  This allows a vector space to be defined on the
``basis vectors'' of $q$ and $r$.  The vector equation of equality
between the $U(1)_{\rm a}$ $D$-term parametrization and the other
parametrizations can be written as
\bea
(\vec v_{\rm a}-\vec v_Y-\vec v_{\rm X})\cdot \vec \lambda_i =0
\eea
where $\vec v_{\rm a}=(q,r)D_{\rm a}$, $\vec v_Y=(q_Y,r_Y)D_Y$, $\vec
v_{\rm X}=(q_{\rm X},r_{\rm X})D_{\rm X}$, and
$\vec\lambda_i=(a_i,b_i)$ such that $Q_{\rm a}^i = a_iq +b_ir$. The
solution is therefore independent of $i$ (i.e., independent of particle
type), and given by Eqs.~(\ref{Damap}) and (\ref{qmap}).

As a short digression, we remark that the $D$-term solution of
Ref.~\cite{Katz:1999uw}, on the other hand, cannot be mapped using our
analysis because the symmetries considered are not special cases of
$U(1)_{\rm a}$. While that study cleverly exploited the higher-order
non-decoupling of $D$-term contributions to mass, the examples were not
simply consistent with gauge coupling unification.  The features of the
symmetries that disrupt gauge coupling unification in that example are
also what make it incapable of being described as a special case of
$U(1)_{\rm a}$.

In principle, one could tell the difference between a general
$U(1)_{\rm a}$ $D$-term contribution to the supersymmetry masses and
contributions that arise from the combination of $D_Y$ and $D_{B-L}$
(or $D_{\rm J}$). The typical map of the latter to the former
(Eqs.~\ref{Damap} and~\ref{qmap}) generates an irrational $q$ charge. 
This means the ratio of $U(1)_{\rm a}$ charges among the superpartners
would be irrational, which is not expected for a sensible $U(1)$
theory.  Unfortunately, we see no reasonable way these possibilities
could be distinguished experimentally via superpartner measurements
alone.

To add to the potential complexity of interpreting experiment, kinetic
mixing of the form $\int d^2\theta \, {\cal W}_Y{\cal W}_{\rm a}$ can
also significantly change the effective charges, even if there is
only one extra $U(1)_{\rm a}$ symmetry in the low-energy theory.  If we
apply the methods of Ref.~\cite{Dienes:1996zr} to the present case of a
not-so-hidden extra $U(1)_{\rm a}$, we find that a hypercharge-mixed
$D$-term contribution arises.  Its effects can be parametrized as,
\beq
D_{\rm a}^{\rm eff} \longrightarrow (1+\chi r_Y)D_{\rm a}
\eeq
and
\beq
q^{\rm eff}\longrightarrow \frac{q+\chi q_Y}{1+\chi r_Y}.
\eeq
This is equivalent to the generation of the hypercharge $D$-term of
Ref.~\cite{Arkani-Hamed:2000xj}, except we are mixing it with a
$U(1)_{\rm a}$ symmetry that is not necessarily the special case
$U(1)_{B-L}$.  We then map it back into a single $D$-term of $U(1)_{\rm
a}$ again, albeit with a new charge $q^{\rm eff}$ which is likely to be
irrational.

In the preceeding, we have shown that we can map the entire space of
interesting $D$-term induced solutions to the tachyonic slepton problem
that do not disrupt supersymmetric gauge coupling unification into
the two parameters $q$ and $D_{\rm a}$ of the ancillary $U(1)_{\rm a}$. 
Furthermore, this general result is especially satisfying, because
it allows us to prove several interesting features of the superpartner
spectrum independent of any specific model details. For example, it
manifestly demonstrates that there is a definite sign of the $D$-term
mass correction to each superpartner in this general approach. It also
allowed us to prove the bound of all superpartners as expressed in
Eq.~(\ref{masstheorem}) within any specific model consistent with our
starting axiom that gauge coupling unification is not disrupted.
A $D$-term solution of this motivated type leads to these unambiguous
propositions regarding the superpartner spectrum and is therefore
subject to experimental refutation.


\section*{Acknowledgements}
BM is supported by the U.S. Department of Energy  under contract
W-31-109-Eng-38.
JDW is supported in part by the U.S. Department of Energy
and the Alfred P. Sloan Foundation.



\begin{thebibliography}{99}

\bibitem{Randall:1998uk}
L.~Randall and R.~Sundrum,
``Out of this world supersymmetry breaking,''
Nucl.\ Phys.\ B {\bf 557}, 79 (1999)
[hep-th/9810155].

\bibitem{Giudice:1998xp}
G.~F.~Giudice, M.~A.~Luty, H.~Murayama and R.~Rattazzi,
``Gaugino mass without singlets,''
JHEP {\bf 9812}, 027 (1998)
[hep-ph/9810442].

\bibitem{Pomarol:1999ie}
A.~Pomarol and R.~Rattazzi,
``Sparticle masses from the superconformal anomaly,''
JHEP {\bf 9905}, 013 (1999)
[hep-ph/9903448].

\bibitem{Bagger:1999rd}
J.~A.~Bagger, T.~Moroi and E.~Poppitz,
``Anomaly mediation in supergravity theories,''
JHEP {\bf 0004}, 009 (2000)
[hep-th/9911029].

\bibitem{Luty:2001zv}
M.~Luty and R.~Sundrum,
``Anomaly mediated supersymmetry breaking in four dimensions, naturally,''
hep-th/0111231.

\bibitem{Gherghetta:1999sw}
T.~Gherghetta, G.~F.~Giudice and J.~D.~Wells,
``Phenomenological consequences of supersymmetry with anomaly-induced masses,''
Nucl.\ Phys.\ B {\bf 559}, 27 (1999)
[hep-ph/9904378].

\bibitem{Katz:1999uw}
E.~Katz, Y.~Shadmi and Y.~Shirman,
``Heavy thresholds, slepton masses and the $\mu$ term in anomaly mediated  supersymmetry breaking,''
JHEP {\bf 9908}, 015 (1999)
[hep-ph/9906296].

\bibitem{Chacko:1999am}
Z.~Chacko, M.~A.~Luty, I.~Maksymyk and E.~Ponton,
``Realistic anomaly-mediated supersymmetry breaking,''
JHEP {\bf 0004}, 001 (2000)
[hep-ph/9905390].

\bibitem{Carena:2000ad}
M.~Carena, K.~Huitu and T.~Kobayashi,
``RG-invariant sum rule in a generalization of anomaly mediated SUSY breaking models,''
Nucl.\ Phys.\ B {\bf 592}, 164 (2001)
[hep-ph/0003187].

\bibitem{Jack:2000cd}
I.~Jack and D.~R.~Jones,
``Fayet-Iliopoulos $D$-terms and anomaly mediated supersymmetry breaking,''
Phys.\ Lett.\ B {\bf 482}, 167 (2000)
[hep-ph/0003081].
D.~R.~Jones,
``Anomaly mediated supersymmetry breaking, $D$-terms and $R$-symmetry,''
in {\it Proc. of the 5th International Symposium on Radiative Corrections} (RADCOR 2000) ed. Howard E. Haber,
arXiv:hep-ph/0101159.

\bibitem{Chacko:2000wq}
Z.~Chacko, M.~A.~Luty, E.~Ponton, Y.~Shadmi and Y.~Shirman,
``The GUT scale and superpartner masses from anomaly mediated supersymmetry breaking,''
Phys.\ Rev.\ D {\bf 64}, 055009 (2001)
[hep-ph/0006047].

\bibitem{Allanach:2000gu}
B.~C.~Allanach and A.~Dedes,
``$R$-parity violating anomaly mediated supersymmetry breaking,''
JHEP {\bf 0006}, 017 (2000)
[hep-ph/0003222].

\bibitem{Kaplan:2000jz}
D.~E.~Kaplan and G.~D.~Kribs,
``Gaugino-assisted anomaly mediation,''
JHEP {\bf 0009}, 048 (2000)
[hep-ph/0009195].

\bibitem{Arkani-Hamed:2000xj}
N.~Arkani-Hamed, D.~E.~Kaplan, H.~Murayama and Y.~Nomura,
``Viable ultraviolet-insensitive supersymmetry breaking,''
JHEP {\bf 0102}, 041 (2001)
[hep-ph/0012103];
R.~Harnik, H.~Murayama and A.~Pierce,
``Purely four-dimensional viable anomaly mediation,''
JHEP {\bf 0208}, 034 (2002)
[hep-ph/0204122].

\bibitem{Okada:2002mv}
N.~Okada,
``Positively-deflected anomaly mediation,''
Phys.\ Rev.\ D {\bf 65}, 115009 (2002)
[hep-ph/0202219].

\bibitem{Nelson:2002sa}
A.~E.~Nelson and N.~T.~Weiner,
``Extended anomaly mediation and new physics at 10-TeV,''
hep-ph/0210288.

\bibitem{Jack:2003qg}
I.~Jack and D.~R.~Jones,
``Yukawa Textures and Anomaly Mediated Supersymmetry Breaking,''
hep-ph/0301163.

\bibitem{Chamseddine:1995rs}
A.~H.~Chamseddine and H.~K.~Dreiner,
``Anomaly-free gauged $U(1)'$ in local supersymmetry and baryon number violation,''
Nucl.\ Phys.\ B {\bf 447}, 195 (1995)
[hep-ph/9503454].

\bibitem{Dienes:1996zr}
K.~R.~Dienes, C.~F.~Kolda and J.~March-Russell,
``Kinetic mixing and the supersymmetric gauge hierarchy,''
Nucl.\ Phys.\ B {\bf 492}, 104 (1997)
[hep-ph/9610479].

\end{thebibliography}
\end{document}